\documentclass[10pt,a4paper,twoside]{article}
\usepackage{latexsym,amssymb}
\usepackage{hyperref}
\usepackage{listings}
\usepackage{xcolor}
\usepackage{url}

\usepackage{csquotes}

\usepackage{graphicx}
\usepackage{pgfplots}
\usepackage{soul}
\usepackage{tikz}

\usepgfplotslibrary{units}
\usetikzlibrary{arrows, positioning}
\usetikzlibrary{shapes}

\definecolor{codegreen}{rgb}{0,0.6,0}
\definecolor{codegray}{rgb}{0.5,0.5,0.5}
\definecolor{codepurple}{rgb}{0.58,0,0.82}
\definecolor{backcolour}{rgb}{0.95,0.95,0.92}

\lstdefinestyle{mystyle}{
    language=C++,
    backgroundcolor=\color{backcolour},   
    commentstyle=\color{codegreen},
    keywordstyle=\color{magenta},
    numberstyle=\tiny\color{codegray},
    stringstyle=\color{codepurple},
    basicstyle=\ttfamily\footnotesize,
    breakatwhitespace=false,         
    breaklines=true,                 
    captionpos=b,                    
    keepspaces=true,                 
    numbers=left,                    
    numbersep=5pt,                  
    showspaces=false,                
    showstringspaces=false,
    showtabs=false,                  
    tabsize=2,
    xleftmargin=1cm,
    aboveskip=15pt,
    belowskip=10pt
}

\begin{document}
\begin{center}{\Large\bf
Detecting lifetime errors of\\\texttt{std::string\_view} objects in C++
}\end{center}
\begin{center}{\large\bf\noindent
R\'eka Kov\'acs, G\'abor Horv\'ath, and Zolt\'an Porkol\'ab
}\\[2mm]
Department of Programming Languages and Compilers\\
Faculty of Informatics, E\"otv\"os Lor\'and University\\
Budapest, Hungary
\\[1mm]\texttt{
kovacs.reka@inf.elte.hu\\
xazax@caesar.elte.hu\\
gsd@inf.elte.hu
}\end{center}
\vspace*{7mm}
\begin{abstract}
\noindent \texttt{std::string\_view} is a reference-like data structure in the C++ Standard 
Template Library (STL) that enables fast
and cheap processing of read-only strings. Due to its wide applicability and 
performance enhancing power, \texttt{std::string\_}\texttt{view} has been very popular
since its introduction in the C++17 standard. However, its careless use can lead to 
serious memory management bugs. As the lifetime of a \texttt{std::string\_view} is
not tied to the lifetime of the referenced string in any way,
it is the user's responsibility to ensure that the view is only used while
the viewed string is live and its buffer is not reallocated.

This paper describes a static analysis tool that finds programming
errors caused by the incorrect use of \texttt{std::string\_view}. Our work
included modeling \texttt{std::string\_view} operations in the analysis, 
defining steps to detect lifetime errors, constructing user-friendly diagnostic 
messages, and performing an evaluation of the checker.
\end{abstract}

\section{Introduction}
\label{intro}

C++ is a popular language of choice for performance critical programs.
One important reason for this is its backwards compatibility with the C language~\cite{stroustrup2002c},
allowing developers to build on existing, trusted systems.
However, combining language elements from C and C++ naively can lead to less
readable or less efficient programs (see Section~\ref{view}). To counter this,
many recent efforts from the standards committee center around new solutions
that make popular coding patterns either simpler or more efficient (or both).

One recent addition to the official set of libraries for the language (to the
Standard Template Library, or STL) is the \texttt{std::string\_view} 
class\footnote{The original template class, \texttt{std::basic\_string\_view}, has 
different specializations for different character types. For the purposes of
this paper, we will use the version for the plain \texttt{char} type,
which is simply called \texttt{std::string\_view}. The same applies to
\texttt{std::basic\_string} and \texttt{std::string}.}.
It is a lightweight, pointer-like data structure that enables fast and cheap
processing of read-only strings. However, its incorrect use
can lead to heap-use-after-free and stack-use-after-return 
errors~\cite{Kostya}. 
%The source of the problem is that a
%\texttt{std::string\_view} created for a certain \texttt{std::string} object
%can only be used as long as the string is live and unchanged.  %% GABOR: ha egy karaktert valtoztatok csak meg pl, az nem invalidalja a string-view-t. Jo az unchanged?
The source of the problem is that after the string is destroyed or reallocated,
the view points to freed memory. Such a view is said to be \emph{dangling}.
Reading through a dangling pointer is an error.

To avoid memory errors as early as possible, experts recommend using static
analysis tools during development~\cite{business}. Most of them run in text editors
and IDEs with as little effort as clicking a button, and some of them are able 
to diagnose sophisticated programming errors. One of these tools is the
Clang Static Analyzer~\cite{kremenek2008finding}, an open-source analyzer that 
comes packaged together with one of the most widely used C++ compilers, 
Clang~\cite{lattner2008llvm}. The Clang Static Analyzer is symbolic execution 
engine~\cite{king1976symbolic} that performs a path-sensitive analysis
of a program by tracking program states and building a model of possible
execution paths. It is actively used and maintained by tech giant Apple, and
promoted as an important feature of their code editor, Xcode~\cite{AppleAnalyze}.
It is also available in Microsoft's popular Visual Studio IDE through the
Clang Power Tools extension~\cite{PowerTools}.

One of the great strengths of the Clang Static Analyzer is its extensibility - 
developers can use the infrastructure of the analyzer to find new classes of bugs 
by writing a new module called a \textit{checker}. A recently introduced 
error-finding module called \texttt{cplusplus.InnerPointer}~\cite{kovacs2019detecting} 
tackled a very similar issue to the \texttt{std::string\_view} problem
outlined above: dangling raw pointers to \texttt{std::string} buffers. Raw 
pointers pointing to the inner buffer of a \texttt{std::string} are often used 
for cheap, no-copy string processing similarly to \texttt{std::string\_view}s,
and they can cause the same kind of memory errors.

Although raw pointers to \texttt{std::string} buffers are very similar in spirit to 
\texttt{std::string\_view}s, they are processed differently by the analyzer.
While raw pointers are essential parts of the language and their handling is 
hard-wired into the engine, \texttt{std::string\_view}s are complicated data 
structures that it does not understand, and thus their 
operations need to be recognized and explicitly modeled during the analysis.

This paper describes the details of a static bug-finding module that finds 
programming errors caused by the incorrect use of \texttt{std::string\_view}. 
For this, we needed to teach the analyzer tool to understand 
\texttt{std::string\_view} operations, define steps to detect lifetime errors, 
construct user-friendly diagnostic messages, and perform an evaluation of
the checker.

%%%%%%
\section{Related work}
\label{related}

Memory safety errors have been the greatest problem for C family languages
since their inception. They can cause crashes, unexpected behavior, and can even
be exploited for malicious security attacks. In C++, creators have long encouraged 
programming patterns that make it more difficult to commit such errors,
e.g. the famous RAII pattern~\cite{stroustrup1994design}.
% FIXME: more of these?
However, it appears that they are still prevalent to the present day.
Last year, Microsoft security engineer Matt Miller revealed that out of all
the security vulnerabilities fixed at the company, 70\% of them still tackle
memory safety issues~\cite{Microsoft}. Google reported the same ratio for
the Chrome project~\cite{Chrome}.

Language contributors see the problem and are constantly trying to come up with
new solutions. One recent effort comes from the chair of the C++ standards
committee, Herb Sutter. In his paper~\cite{Herb:2018}, he proposed an automatic 
categorization of C++ classes by compilers~\cite{horvath2019categorization}, based on their lifetime related 
characteristics. \emph{Lifetime profiles} can distinguish owner-like
(e.g. \texttt{std::string}) and pointer-like (e.g. \texttt{std::string\_view}) 
objects in the program, and issue warnings to prevent memory errors at compilation
time. Lifetime profiles are under implementation in two of the three major C++
compilers, MSVC~\cite{MSVCLifetime} and Clang~\cite{HorvathGehre}.

The main drawback of the lifetime profiles introduced in compilers is that
they rely on herustics, and therefore require the program to be written in a
specific style to produce good results. Unfortunately, many (especially older)
codebases do not satisfy this requirement, and for them lifetime warnings are
not yet practical.

Projects like those can, however, benefit from static bug finding tools such as
the Clang Static Analyzer. Because it performs a more powerful, more expensive 
analysis than the compiler, it can find memory safety problems without strong
style prerequisites. Its dangling string pointer detecting module mentioned in the
introduction (\texttt{cplusplus.InnerPointer}) has found use-after-free errors in 
popular open-source libraries such as GPGME, Facebook's RocksDB, and the Clang 
toolchain's debugger, LLDB~\cite{kovacs2019detecting}.

% FIXME: mas?

%%%%%%
\section{Symbolic execution}
\label{symbolicexecution}

This section introduces symbolic execution~\cite{king1976symbolic} in depth, 
concentrating on methods used by the Clang Static Analyzer.
Its main feat is assigning a \emph{symbol} to represent each unknown (but fixed)
value in the program, collecting information about their possible real values 
at each location (represented by a logical formula called a 
\emph{path constraint}). At each branching statement, path constraints 
are used to calculate which branches can be possibly 
taken in real execution runs. Any branch that is deemed infeasible is abandoned
by the analysis - an important strategy to avoid bogus error reports.
Feasible branches are classified according to their unique constraints, and
pursued separately. In contrast to \emph{flow-sensitive} analyses, 
\emph{path-sensitive} analysis is more precise, meaning that it emits fewer
bogus reports.

%This technique allows the analyzer to find errors that
%only exist in one particular execution path of the program, making the analysis 
%\emph{path-sensitive}. In contrast, \emph{flow-sensitive} analyses that
%produce compiler warnings report errors found on all execution paths.

The Clang Static Analyzer combines classic symbolic execution with the graph 
reachability based data flow analysis ideas described in~\cite{reps1995precise}.
During the symbolic execution process, its core engine
builds a data structure that records the knowledge collected about the program
at each step, called an \emph{exploded graph}. Each node of the exploded graph is a
\emph{symbolic program state}, corresponding to a set of real 
program states. However, a standard simulation of the program
will not find programming errors by itself. The engine was therefore designed 
to work together with a set of plug-in bug-finding modules called
\emph{checkers}. Checkers actively participate in the simulation process,
creating new nodes, and storing extra bug-specific information in the exploded
graph. They also define the rules that specify an erroneous state, in which
case the analysis is aborted on that path and the error is reported to the user.

The Clang Static Analyzer performs context-sensitive interprocedural analysis.
In the beginning, the analyzer picks a \emph{top-level function} and starts to
interpret it without any calling context. When there is a call to a function with 
a known body, the analyzer \emph{inlines} the call, i.e. continues the analysis
inside the callee, preserving all the information known at the call site. If the 
function body is not available, the call will be evaluated \emph{conservatively}.
After the analysis is done, the engine picks a new top-level function which has
not been visited (inlined) during the analysis of earlier top-level functions.

\begin{figure*}[h!]
\centering
\noindent\begin{minipage}{.35\textwidth}
\centering
\begin{lstlisting}[frame=tlrb, xleftmargin=1cm, basicstyle=\ttfamily\footnotesize]{Name}
void g(int b, int &x) {
  if (b)
    x = b + 1;
  else
    x = 42;
}
\end{lstlisting}
\end{minipage}\hfill
\begin{minipage}{.64\textwidth}
\centering
\begin{tikzpicture}[scale=0.8, transform shape]
\draw  (-2,2.5) rectangle (1,1) node[pos=.5, rectangle split,rectangle split parts=3] {b: $\$b$, x: $\$x$\nodepart{second} \$b : [IMIN, IMAX]\nodepart{third} \$x : [IMIN, IMAX]};
\draw  (-4.5,0) rectangle (-1.5,-1.5) node[pos=.5, rectangle split,rectangle split parts=3] {b: $\$b$, x: $\$x$\nodepart{second} \$b : [0, 0]\nodepart{third} \$x : [IMIN, IMAX]};
\draw  (-4.5,-2.5) rectangle (-1.5,-3.5) node[pos=.5, rectangle split,rectangle split parts=2] {b: $\$b$, x: $42$\nodepart{second} \$b : [0, 0]};
\draw  (0.5,0) rectangle (5,-1.5) node[pos=.5, rectangle split,rectangle split parts=3] {b: $\$b$, x: $\$x$\nodepart{second} \$b : [IMIN, -1]$\ \cup\ $[1, IMAX]\nodepart{third} \$x : [IMIN, IMAX]};
\draw  (0.5,-2.5) rectangle (5,-3.5) node[pos=.5, rectangle split,rectangle split parts=2] {b: $\$b$, x: $\$b+1$\nodepart{second} \$b : [IMIN, -1]$\ \cup\ $[1, IMAX]};
\draw [->](-0.5,1) node (v1) {} -- (-3,0);
\draw [->](v1) -- (2.5,0);
\draw [->](-3,-1.5) -- (-3,-2.5);
\draw [->](2.5,-1.5) -- (2.5,-2.5);
\end{tikzpicture}
\end{minipage}
\caption{A simplified version of the exploded graph built during the symbolic execution of a simple function. The right-hand side of the graph represents the \textit{true} branch of the \texttt{if} statement, while the left-hand side describes the \textit{false} branch.}
\label{fig:exploded}
\end{figure*}
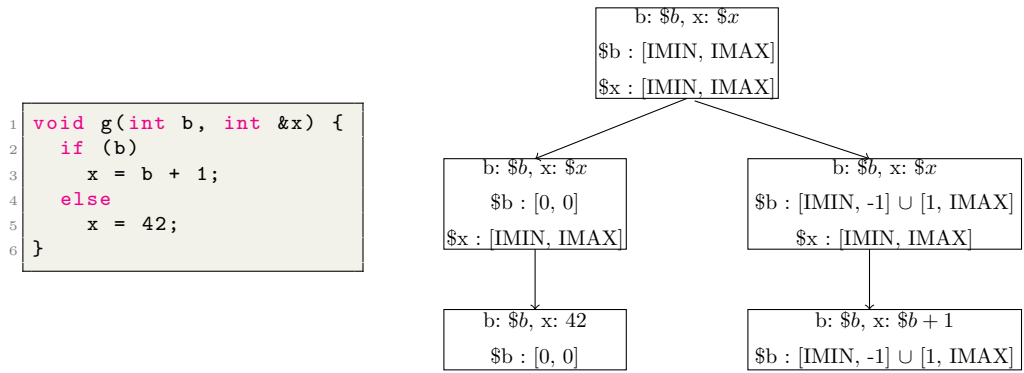

An example analysis can be seen along with its simplified exploded graph in 
Figure~\ref{fig:exploded}. The first line of each program state box is the 
\textit{store}, which maps memory locations to symbolic expressions. The other
lines represent path constraints over symbols collected during exploration.
We omitted any remaining components for brevity.

% FIXME: graf dobozok tul kicsik.

In this example, function \texttt{g} has two execution paths. Since the values of 
\texttt{b} and \texttt{x} are initially unknown, these values are represented by 
the corresponding symbols \texttt{\$b} and \texttt{\$x}.
As the analysis continues, on one of the execution paths the
value of \texttt{b} is known to be zero. Later on this path we discover that
the value of x is the constant \texttt{42}. The symbol \texttt{\$x} is no longer
needed on this path. On the second path, the value of \texttt{b} can be anything
but zero. On this path, we also discover that the value of \texttt{x} is one larger
than the original value of \texttt{b}. The symbol \texttt{\$x} is no longer needed
on any of the paths, it can be garbage collected.

%%%%%%
\section{The \texttt{std::string\_view} class}
\label{view}

One of the greatest selling points of the C++ programming language is its
backwards compatibility with older systems written in C, while being a 
modern, high-level language capable of multiple paradigms and programming styles.
The need to keep it compatible has placed a great burden on the language over
the years, and made it increasingly complicated. Recent efforts~\cite{Herb:2019}
try to introduce new, ingenious solutions to the language in order to win back
some of the simplicity that had been lost.

One notable effort to win back simplicity and at the same time improve the performance
of string-processing programs was the introduction of the
\texttt{std::string\_view}~\cite{Yasskin:2013}
data structure to the C++ Standard Template Library (often abbreviated as STL)
in 2017.

Beforehand, C++ programmers could choose to handle strings in one of two ways. 
They could either pass around a pointer to a read-only, C-like string literal 
(a \texttt{const char *} type object), or create a C++ \texttt{std::string} object.
C-like string literals are generally managed by the compiler, ensuring that
the same character sequence is only stored once in the program. 
\texttt{std::string}s, on the other hand, store a modifiable, exclusively owned
copy of their contents, managed by the programmer. As \texttt{std::string}
objects can be fairly large, they are normally passed around functions in a 
\emph{reference to constant string} format (\texttt{const std::string \&}), 
to avoid unnecessary argument copies.

The combination of these two string formats often led to a decline in
the quality of a program. This can best be illustrated by 
examples originating from~\cite{Chastain:2012}.

\begin{lstlisting}[caption={C convention for taking a string as a parameter.},label={lst:c-conv}]
void TakesCharStar(const char *s);

void HasString(const std::string &s) {
  TakesCharStar(s.c_str()); // explicit conversion
}
\end{lstlisting}

\noindent In the example in Listing~\ref{lst:c-conv},
the C++ string object needs to be explicitly converted to \texttt{const char *}
type by the \texttt{c\_str} method, making the program more difficult to read.

\begin{lstlisting}[caption={Old standard C++ convention
(before C++17) for taking a string as a parameter.},
label={lst:old-cpp-conv}]
void TakesString(const std::string &s);

void HasCharStar(const char *s) {
  TakesString(s);           // compiler will make a copy
}
\end{lstlisting}

\noindent Although the example in Listing~\ref{lst:old-cpp-conv} is 
comfortable to read, it leaves precious performance on the table. When a C++ string is
created from a C-like string literal, the whole character sequence is copied.
This is necessary because of the semantics of \texttt{std::string} - it needs
to be able to own and modify the string.

\texttt{std::string\_view} is a new C++ class in the STL that aims to solve both 
of these problems. By essentially being a \texttt{const char *} wrapped in a C++
object, it combines the lightweightness of a C pointer and the flexibility of
a C++ class. This also means that \texttt{std::string\_view}s do not store a copy
of the viewed character sequence, they merely reference it.

To improve the situation in the first example, the \texttt{std::string} class
has been modified in the C++17 standard to convert to \texttt{std::string\_view} 
implicitly. Furthermore, to avoid the spurious copy in the second example, the 
semantics of \texttt{std::string\_view} have been defined so that its construction
from a string literal only saves a pointer to the string, and does not copy
the whole character sequence.

Using these new tools provided by the C++17 standard, string processing in
modern codebases looks like Listings \ref{lst:new-cpp-conv-1} and 
\ref{lst:new-cpp-conv-2} below.

\begin{lstlisting}[caption={New standard C++ convention 
(after C++17) for taking a string as a parameter.},label={lst:new-cpp-conv-1}]
void TakesStringView(std::string_view s);

void HasString(const std::string &s) {
  TakesStringView(s);   // no explicit conversion!
}
\end{lstlisting}

\begin{lstlisting}[caption={New standard C++ convention 
(after C++17) for taking a string as a parameter.},label={lst:new-cpp-conv-2}]
void TakesStringView(std::string_view s);

void HasCharStar(const char *s) {
  TakesStringView(s);   // no copy; efficient!
}
\end{lstlisting}

What is interesting about \texttt{std::string\_view} is that various 
implementations of the same concept had existed long before its
standardization, including the \texttt{StringPiece}~\cite{AbseilStrings} class in the Google
libraries (later renamed to \texttt{absl::string\_} \texttt{view}),
\texttt{String\_Ref}~\cite{Marshall:2012} in the Boost library (commonly considered the official
playground of potential standard C++ features),
or \texttt{StringRef}~\cite{StringRef} in LLVM (a collection of libraries widely used to
write compilers). % https://llvm.org/doxygen/classllvm_1_1StringRef.html
However, based on a quick search on GitHub, countless smaller C++ projects have
implemented their own versions as well. It seems that the success of the idea
hinders the adoption of the standard \texttt{std::string\_view} data structure - 
many projects already implemented a version of it, and are slow in updating 
their codebase, especially as they need to adopt the whole C++17 standard.
On the bright side, we saw many project owners express their intention to do
so in the future.

\section{Detecting \texttt{std::string\_view} errors}
\label{detecting-errors}

%After the previous section, it is reasonable to expect that 
%\texttt{std::string\_view} will become widely adopted in the coming years.
%In that case, developers will need to look out for new kinds of potential
%memory errors in their code. 

The internal representation of a \texttt{std::string\_view} is essentially
a pointer and a length (or two pointers - one to the beginning, one to the end 
of the string), wrapped in a C++ object. It is a snapshot of the string
from which it was created, meant to be used as long as the string is live
and its buffer is not reallocated. Whenever these characteristics of the string 
change, all previously created \texttt{std::string\_view}s are considered 
\emph{invalid}.

The problem is that the view object is not associated with its string in any way,
only in the programmer's mind. If the programmer forgets to check whether the 
view is still usable, the compiler will not be able to do it for them, and the 
situation may lead to use-after-free errors.

One example of such an error can be seen in Listing~\ref{lst:stack-use-after-return}. 
Here, the caller will receive a view that references a local string that was 
destroyed at the end of the called function.

\begin{lstlisting}[caption={An example of a stack-use-after-return error.},label={lst:stack-use-after-return}]
std::string_view stackUseAfterReturn(bool cond) {
  std::string s("MaCS");
  std::string_view v(s);
  if (cond)
    return v; // View points into the deallocated string!
}
\end{lstlisting}

Another example can be seen in Listing~\ref{lst:heap-use-after-free}.
In this case, the view points into a string that is destroyed at the end of
its scope, at the closing bracket on Line 6. Any consequent use of the view
is a use-after-free error.

\begin{lstlisting}[caption={An example of a heap-use-after-free error.},label={lst:heap-use-after-free}]
char heapUseAfterFree(bool cond) {
  std::string_view v;
  {
    std::string s("MaCS");
    v = s;
  }
  if (cond)
    return v[0]; // View points into the deallocated string!
}
\end{lstlisting}

These and similar memory bugs can lead to crashes, hard-to-debug unexpected
behaviors, and can even be exploited for malicious security attacks. 
In this section, we describe a set of modules for the Clang Static Analyzer
that work together to report lifetime issues to the programmer early in
the development process.

\subsection{Overview}
\label{overview}

Following the timeline of symbolic execution, we need to complete the following
basic steps to find lifetime-related \texttt{std::string\_view} bugs:

\begin{enumerate}
\item Recognize that a view is created. Save an association between the view
and the string it references.

\item If the view is modified to point to another string, update the association.

\item Recognize if the string is modified or destroyed, as described by the
C++ standard. Mark all views referencing that string \emph{released}.

\item Recognize statements that indicate a use of a \emph{released} view. Emit a
warning with a descriptive error message.

\item Augment the bug path with additional explaining notes.
\end{enumerate}

For the third point, we did not have to write the string pointer invalidation
logic, as it had already been implemented in another module called 
\texttt{cplusplus.InnerPointer}~\cite{kovacs2019detecting}. We could completely
reuse that logic and only had to add the action of marking all the views referencing
an invalidated string released.

\subsubsection{Data structures}

We will often reference the following data structures created for the
purposes of our analysis:

\begin{itemize}
    \item \textbf{\texttt{ViewRegions} map.} This data structure maps each string 
    object to a set of view objects that reference it. (Both kinds of objects are 
    represented as pointers to memory regions.) It is maintained so that it stores 
    up-to-date information at each program point. If a string is modified or
    destroyed, this map is used to mark all associated views released. At that
    point, the corresponding entry is deleted.
    
    \item \textbf{\texttt{CastSymbols} map.} This is an auxiliary map only needed
    for technical reasons. It maps each string to a set of symbols that are
    generated as the result of the string-to-view conversions happening
    whenever a view is created from a string directly. These cast symbols are
    needed to bridge the temporal gap between the disappearance of the string object
    and the appearance of the created view object in the exploded graph. At the
    point where the view object is finally created by the analysis engine, the 
    referenced string is looked up in this map, and the \emph{(string, view)} pair is 
    added to \texttt{ViewRegions}.

    \item \textbf{\texttt{ReleasedViews} set.} When a string object is modified
    or destroyed, all the view objects referencing it (as stored by the
    \texttt{ViewRegions} map) are loaded into this set. We issue a bug report
    whenever a view contained in this set is used in the program.
\end{itemize}

\subsection{Preliminary string modeling}
\label{string-modeling}

Before we began executing the plan outlined in Section~\ref{overview}, 
we made an addition that is not strictly needed for the bug detection process.

\texttt{c\_str} and \texttt{data} are two \texttt{std::string} methods 
that allow the programmer to obtain direct access to the inner buffer of a 
string, by returning a raw pointer pointing to it. Standard library function calls often
cannot be inlined to the program, and are instead evaluated conservatively,
treating them like a black box. In that case, the analyzer will safely assume 
that different subsequent calls of the same unknown function return a different
result. However, we are familiar with the semantics of these particular standard 
methods, and know that they return the same pointer on every call as long as
they are called on the same string, and the buffer of that string has not been 
reallocated.

To incorporate this extra domain specific information to the analysis,
we take over the task of managing symbols for these calls from the engine.
When a pointer to the inner buffer is first requested for a given string object, 
we create a new, unique symbol to represent it, and bind it to the call expression 
in the \emph{environment} part of the program state. This ensures that the new symbol
will be automatically propagated by the engine as the result of the call expression.
We record the \emph{(string, buffer symbol)} pair in a checker-specific map in the 
program state, allowing us to return the same unique symbol for every 
\texttt{c\_str} and texttt{data} call on the same, not reallocated string object. 
This includes the case when the string is accessed through the 
\texttt{std::string\_view::data} method. Whenever the string becomes re- or
deallocated, the \emph{(string, buffer symbol)} entry is deleted, and the next
\texttt{c\_str}/\texttt{data} call generates a new symbol again.

We implemented these string modeling additions in a new module that we called
\texttt{cplusplus.} \texttt{StringModeling}. With them, we not only made our 
analysis more precise, but also made it simpler and faster, as we no longer 
have to follow a set of different buffer symbols for the same string.

\subsection{View modeling}
\label{view-modeling}

The first event that we need to recognize during the execution of the program
is the creation of a \texttt{std::string\_view} from a \texttt{std::string}.
Views created from C-like string literals are not of interest to us, as they
have static storage duration and therefore do not cause lifetime related issues. 
We do not support views created from arrays and iterators yet.

In the cases that concern us, a view first has to be created from a 
\texttt{std::string} object directly. This \emph{(string, view)} association
can then be transitively propagated when views are created from other views.
We also have to track operations that modify a given view, to ensure that
we always associate the right string object with it.

All the features described in this section were implemented
in a module that we named \texttt{cplusplus.StringViewModeling}.

\subsubsection{Creating a view from a string}
\label{from-string}

This can happen in one of two fundamentally different ways. In the first case,
illustrated by Listing~\ref{lst:view-creation-1}, we create a view by calling
its copy constructor with a \texttt{std::string} argument.
Note that the basic\_string\_view class does not have a constructor that
takes a basic\_string parameter. The copy constructor call works because a
basic\_string\_view conversion operator was added to the basic\_string class
in the C++17 standard. The string is therefore implicitly converted to a view
before the copy.

\begin{lstlisting}[caption={Creating a view from a string by passing it to the
copy constructor.},label={lst:view-creation-1}]
  std::string s("abc");
  std::string_view v = s;
\end{lstlisting}

In the second case, which can be seen in Listing~\ref{lst:view-creation-2},
an empty view is created first by calling the default constructor. The string
is then associated with this view by the copy assignment operator call. The
string is implicitly converted to a view in the same fashion as in the previous
example.

\begin{lstlisting}[caption={Creating a view from a string by first declaring an 
empty view, then passing the string to the copy assignment operator.},
label={lst:view-creation-2}]
  std::string s("abc");
  std::string_view w;
  w = s;
\end{lstlisting}

In order to recognize these events in our checker, we have to "subscribe" to
different AST nodes. Whenever the engine gets to process those nodes, we get a
chance to do our own work, which is to record the association between the
string and the view. We can store this kind of information in the dedicated
data structures that we can declare for our checkers in the program state.

Unfortunately, we found that the string object and the freshly created view object
is never available at the same time at any particular program point, so we cannot
match just one AST event. We therefore took an indirect approach, and chose to store 
the resulting symbol of the string to string\_view conversion as a mediator 
between the two. When the conversion is processed, we store the \emph{(string, 
cast result)} association in a map that we named \texttt{CastSymbols}. 
Later, we look for the moment when the result of the
conversion is bound to the newly created view. In that moment,
the view is finally available to us. We can look up the previously stored
string based on the conversion result symbol, and save the \emph{(string, view)}
pair in our \texttt{ViewRegions} map. We call them regions because we represent
C++ objects as pointers to memory regions.

\subsubsection{Creating a view from another view}

For our bug detection tools, we are also interested in views created from other
views that already store a reference to a std::string. These can be created by
plain copy constructor calls or the \texttt{substr} method. \texttt{substr}
returns a new view object that holds a substring of the original string. In 
Listing~\ref{lst:view-creation-3}, we can see examples of both methods.

\begin{lstlisting}[caption={Creating a view from a an existing one by either 
calling the copy constructor or calling the substr method.},
label={lst:view-creation-3}]
  std::string s("abc");
  std::string_view u = s;
  
  std::string_view v = u;                // copy constructor
  std::string_view w = u.substr(1, 2);   // substr
\end{lstlisting}

In our checker, we handled these cases by matching the copy constructor and
the \texttt{substr} calls respectively. We looked up the copied view in our
\texttt{ViewRegions} map to obtain the string associated with it. We then
recorded the association between the string and the newly created view object
in the same map.

\subsubsection{Modifying a view}

Views can be modified to reference a different string than the one they were
created from. This can be done by copy assignment operator calls or the 
\texttt{swap} method.

The copy assignment operator can be called with either a string or a view
argument. The first case entails the conversion mechanism described in
Section~\ref{from-string}. Examples can be seen for both cases
in Listing~\ref{lst:view-mod-1}.

\begin{lstlisting}[caption={Views can be modified by calling the copy assignment
operator with a string or a view argument.}, label={lst:view-mod-1}]
  std::string s1("abc");
  std::string s2("xyz");
  
  std::string_view v = s1;
  std::string_view w = s2;
  
  v = s2;  // copy assignment with a string argument
  v = w;   // copy assignment with a view argument
\end{lstlisting}

To handle this case in our analysis, we first match the assignment operator.
Then, to simulate overwriting, we need to delete the previously held string 
association of the destination view from the \texttt{ViewRegions} map. (If it
had none, then the view is empty and we are actually in Section~\ref{from-string}).
Next, we look up the argument in the \texttt{ViewRegions} map. If we find a 
corresponding string entry, we can simply add an association between the 
destination view and that string. If there is no such entry, then the argument is 
either a view we do not care about (e.g. created from a string literal), or a string. 
In the latter case, we are already past the string-to-view conversion, and
the resulting symbol of the cast has to be in our \texttt{CastSymbols} map
(see Section~\ref{from-string}). We look up the corresponding string object,
and add the destination view to its \texttt{ViewRegions} entry.

The \texttt{swap} method swaps the contents of two views. Its modeling in the
analysis requires swapping the string associations of the two views. An
example can be seen in Listing~\ref{lst:view-mod-2}.

\begin{lstlisting}[caption={The swap method swaps the contents of two views.}, 
label={lst:view-mod-2}]
  std::string s1("abc");
  std::string s2("xyz");
  
  std::string_view v = s1;
  std::string_view w = s2;
  
  v.swap(w);  // v now references s2, w references s1
\end{lstlisting}

\subsection{String buffer invalidation}

A string\_view essentially stores a raw pointer pointing to the inner buffer of
a string. The conditions under which such pointers are invalidated are described
in the C++ standard. Quoting the \S 21.3.2.1  [string.require] section from its
recent draft N4861~\cite{StdDraft}:

\begin{displayquote}
"References, pointers, and iterators referring to the elements of a
\texttt{basic\_string} sequence may be invalidated by the following uses of that
\texttt{basic\_string} object:
 
-- Passing as an argument to any standard library function taking a reference to 
non-const \texttt{basic\_string} as an argument.

-- Calling non-const member functions, except \texttt{operator[]}, \texttt{at},
\texttt{front}, \texttt{back}, \texttt{begin}, \texttt{rbegin}, \texttt{end}, 
and \texttt{rend}."
\end{displayquote}

Fortunately, these conditions had already been implemented in an existing 
module called \texttt{cplusplus.InnerPointer}~\cite{kovacs2019detecting}
that we briefly mentioned in Section~\ref{intro} and Section~\ref{related}.
This module tracks raw buffer pointers of strings very much
like we track views of strings. When pointers to a string become invalid due
to a condition described in the standard excerpt, it marks all the raw pointers
associated with that string object \emph{released}. The only addition we had to
make is to have it mark \emph{all the views} associated with the string 
\emph{released} as well.

This is done by iterating through the views associated with the given string
(in \texttt{ViewRegions}),
and adding them to another checker-specific data structure that we named 
\texttt{ReleasedViews}. This set is managed by \texttt{cplusplus.StringViewChecker},
which is responsible for reporting the view-related errors to the user 
(see Section~\ref{reporting}). The entry associated with the deallocated string in 
\texttt{ViewRegions} can then be deleted.

\subsection{Reporting the use of released views}
\label{reporting}

Along the interpretation of a program, we may arrive at a point when we have
dangling views in our \texttt{ReleasedViews} set. Arguably, the most important 
task of our analysis is to check if any of them are used, and if they are, emit 
a warning message for the user. We implemented this feature in a dedicated
module called \texttt{cplusplus.StringViewChecker}.

\paragraph{Check return statements}

Returning a dangling view from a function is an easy mistake to make.
In this case, the returned view points into a local string that is destroyed
at the end of the function. The caller will therefore receive a view that
references freed memory. Better warn before that happens.

\begin{lstlisting}[caption={Function returning a dangling view.},
label={lst:return-stmt}]
std::string_view returnDanglingView() {
  std::string s("abc");
  std::string_view v(s);
  s.clear();  // v is invalidated
  return v;   // warning: use-after-free
}
\end{lstlisting}

\paragraph{Check method calls}

Calling methods on a dangling view, apart from its constructors and the copy
assignment operator, is suspicious. In our example in Listing~\ref{lst:method-call},
\texttt{remove\_prefix} shrinks the view by moving its start pointer forward by 1
character. This is not a serious error yet, because we did not attempt to read 
freed memory. However, there is no point in doing this operation unless we plan
to use the modified view. To forego any later uses, we give a warning at this point.

\begin{lstlisting}[caption={Calling a method on a dangling view.},
label={lst:method-call}]
void methodCallOnDanglingView() {
  std::string s("abc");
  std::string_view v(s);
  s = "xyz";           // v is invalidated
  v.remove_prefix(1);  // warning: use after free
}
\end{lstlisting}

We report the error for all \texttt{string\_view} methods except the constructors, 
the assignment operator, and the capacity inquiry methods (\texttt{size}, 
\texttt{length}, \texttt{max\_size}, \texttt{empty}). We feel that requesting
size information about a dangling view may have valid use cases, and it is
important that we avoid unnecessary false positive reports.

\paragraph{Check function arguments}

Passing a dangling view to a library function \emph{by value} or \emph{by const 
reference} is problematic. In the first case, the function will act on a copy of 
the invalid view. In the second case, no copy takes place, but the view has to
be used without a change. Both of these cases imply that
the user plans to use the view in its present, invalid form. Otherwise, the function
would have expected it by non-\texttt{const} reference, so that it could overwrite
it before use.

One popular example of this error, seen in Listing~\ref{lst:func-arg}, is comparing 
a dangling view to a string literal. The corresponding comparison function 
% https://en.cppreference.com/w/cpp/string/basic_string_view/operator_cmp
takes \texttt{string\_view}s by value.

\begin{lstlisting}[caption={Free function taking a dangling view argument
by value.},label={lst:func-arg}]
void functionArgIsDanglingView() {
  std::string s("abc");
  std::string_view v(s);
  s += "xyz";             // v is invalidated
  if (v == "something")   // warning: use after free
    // ...
}
\end{lstlisting}

\paragraph{Check dangling string pointers}

This is a big project that has already been implemented as an interplay between 
\texttt{cplusplus.InnerPointer} and \texttt{unix.MallocChecker}. Our addition
is to find the error if it is committed through a \texttt{string\_view}.

The \texttt{string\_view::data} method allows the user to obtain a raw pointer
pointing to the inner buffer of the associated string. Using this pointer after
the string is modified or destroyed is the same kind of error that we are looking
for, just with pointers instead of views (see Listing~\ref{lst:location}).

\begin{lstlisting}[caption={Function returning a dangling string buffer pointer.},
label={lst:location}]
const char *returnDanglingPointer() {
  std::string s("abc");
  std::string_view v(s);
  s.append("d");     // v is invalidated
  return v.data();   // warning: use-after-free
}
\end{lstlisting}

To empower \texttt{InnerPointer} to catch this error, we need to establish a
connection between the call expression and the buffer symbol
associated with the string. (We described the implementation
of this feature in Section~\ref{string-modeling}.) After detecting the call,
we look up the symbol bound to the view object in the program store. If the
view was created from a \texttt{string}, this symbol should be present in our
\texttt{CastSymbols} map. We loop up the corresponding string in \texttt{CastSymbols},
obtain its buffer symbol from \texttt{InnerPointer}, and bind the buffer symbol
to the call expression in the program state.

Once the \texttt{v.data()} call is associated with the buffer symbol of the string,
and \texttt{InnerPointer} marks the buffer symbol released, \texttt{MallocChecker}
will detect that a released symbol is returned from the function, very much like
we did for views in the previous paragraph.

\subsection{How to use it}
\label{how-to}

As our work is very recent, it is not yet available in the official
Clang repository. However, we have a lot of experience working together
with the Clang Static Analyzer team, and are confident that the upstreaming
process, planned to begin shortly, will be successful. Even more so
because this work is a direct continuation of a previous joint project,
the implementation of the \texttt{cplusplus.InnerPointer} checker.

While the upstreaming process is underway, our working draft is publicly
available as an LLVM fork on GitHub~\cite{workingBranch}. After a standard
Clang build as described in the documentation~\cite{ClangDoc}, one can locate
the built binaries under \texttt{llvm-project/build/bin/}. The easiest
method to run the Clang Static Analyzer is to use the \texttt{scan-build}
command followed by our own build command:

\vspace{-5pt}

\begin{lstlisting}[label={lst:},numbers=none,xleftmargin=0cm,
backgroundcolor={}]
  $ scan-build clang -std=c++17 file.cpp
\end{lstlisting}

\noindent Please note that our analysis only works if the program is compiled in 
C++17 mode. 

\begin{figure}[h!]
\label{report}
\centering
\includegraphics[width=0.5\textwidth]{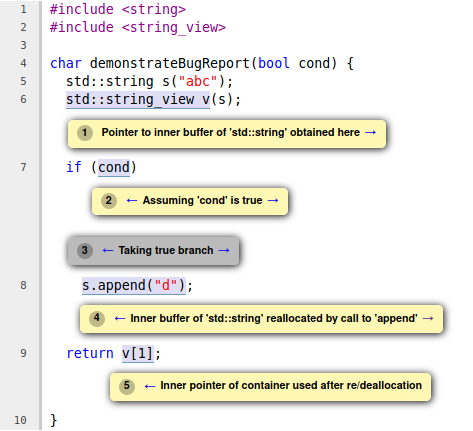}
\caption{A visual representation of the execution path leading to the bug, augmented
with explaining notes, displayed in a browser.}
\end{figure}

Running the \texttt{scan-view} command specified in the instructions displayed 
in the terminal, the user is presented with a summary of the results of the 
analysis run.
Clicking on the \emph{View Report} button next to a bug in the list, the user can browse a visual
representation of the reasons leading to the bug, as displayed in Figure~\ref{report}.

One of the greatest characteristics of the Clang Static Analyzer is that its warnings 
are path-sensitive. This means that users not only receive a warning
on a single line of the file, but a trimmed version of the whole execution 
path leading to the bug, called the \emph{bug path}. The automatically generated
part of the bug path consists of explaining notes placed at important milestones 
in the program (e.g. which branch was taken at a branching statement).
Our checkers also add their own notes that explain bug-specific events along the 
bug path.

Once our analysis is upstreamed to the official LLVM repositories, it will also
become available in Apple's Xcode IDE, and in Microsoft's Visual Studio (through 
the Clang Power Tools extension).

%%%%%%
\section{Evaluation}
\label{eval}

\subsection{Correctness}

The LLVM Compiler Infrastructure Project is a collection of repositories that
contain mission-critical infrastructure software for many tech companies. It is
important that new additions do not break any existing features. Each new
contribution is therefore required to include a set of litmus tests, which are
run by a network of build bots after each commit. Because we plan to commit
our work to the official repository, we made sure to adhere to these standards
and included a collection of regression tests with our checkers.

To test a given bug-finding module of the Clang Static Analyzer, it is not only
important that a warning is given for a test case, but the placement and the
content of each diagnostic message also needs to be right. Special testing
directives (\texttt{expected-warning}, \texttt{expected-note}) are available
for this purpose. The \texttt{@-N} suffix indicates that the note or warning is 
expected to be seen N lines earlier compared to the placement of the directive.
A sample test case can be seen in Listing~\ref{lst:test}. Messages are shortened
for the sake of readability.

\begin{lstlisting}[caption={A sample test case from the regression test suite.},
label={lst:test},basicstyle=\ttfamily\footnotesize]
void CopyAssignUseAfterAutomaticFree() {
  std::string_view V;
  std::string_view W;
  {
    std::string S("abc");
    V = S;
    W = V; // expected-note {{Pointer to inner buffer of 'std::string' ...}}
  } // expected-note {{Inner buffer of 'std::string' deallocated by call ...}}
  W.remove_prefix(1);
  // expected-warning@-1 {{Inner pointer of container used after ...}}
  // expected-note@-2 {{Inner pointer of container used after ...}}
}
\end{lstlisting}

\subsection{Efficacy}

Once the correctness of the checkers is established by isolated regression tests, 
it is important to see how they perform in the real world. For this purpose, it is
customary to run the analysis on several open-source C++ projects, and monitor the
output of the checkers.

Reports emitted (or not emitted) by the checkers can be categorized into four
different classes: true and false positives, and true and false negatives.

Negative reports do not actually exist; they indicate the lack of a report. 
\emph{True negative} means that the lack of a report is appropriate, as the code
under inspection is correct. This is the least useful category, as most of the code
belongs to it. A \emph{false negative} case, on the other hand, should be reported,
but the analysis tool does not find it. These cases can be counted if we know the
project well, and have a list of confirmed errors to look for.

Positive reports are the actual list of reports the tool emits. These can be inspected
by hand to confirm whether they are real or bogus findings. \emph{True positive} reports
are incorrect code parts that are rightfully identified by the analyzer as such.
\emph{False positives}, on the other hand, are code parts reported to be erroneous when
they are actually correct.

Bug finding tools, like the Clang Static Analyzer, are designed to be easy to use and
relatively quick, ideally aiding the developer right in the IDE. They neither intend to 
prove the absence of a class of errors, nor claim that they can find all of the errors 
of some kind. Their aim is to find as many interesting, serious bugs as possible, 
while producing the least amount of bogus reports as possible. It is an intricate balance
that needs to be fine-tuned.

\paragraph{False positives}

The most important requirement towards a bug finding tool is to have a low false positive
rate. Even a completely silent checker is better than one that bombards the user with
useless reports. Inspecting the reports by hand can take hours, even days, in complicated
code bases. It does not take many bogus cases for the user to abandon the tool altogether.

The best way to measure the false positive rate is to run the analysis on many open-source
C++ projects. The criteria for the test project selection are highly dependent on the feature 
to be tested. In our case, we were mostly interested in widely used projects that build with 
the C++17 standard and use \texttt{std::string\_view} extensively.

This task is surprisingly difficult. On the one hand, C++ projects (especially large ones)
often have heterogeneous, unique build systems, which require many dependencies and are
hard to execute. On the other hand, very few projects are so up-to-date as to build with
such a recent standard as C++17. During our exploration on GitHub, we saw that most
projects that perform extensive string processing already implemented their own version
of string\_view years before its appearance in the standard, and they are slow to abandon it.

After a long search for suitable open-source projects, we run our feature on
OpenImageIO~\footnote{https://github.com/OpenImageIO/oiio}, % 
LibreOffice\footnote{https://github.com/LibreOffice/core}, % https://github.com/LibreOffice/core
GPGME\footnote{https://github.com/gpg/gpgme}, and % https://github.com/gpg/gpgme
Ceph\footnote{https://github.com/ceph/ceph}. % https://github.com/ceph/ceph
% tablazat? cite-ok?
The good news is that we did not find any false positive reports from our \texttt{std::string\_view} 
checkers on these projects.

\paragraph{False negatives}

The bad news is that we neither found any true positive reports. Finding the locations
where the analysis missed bugs in such huge, unknown codebases is impossible. What we
can instead do is take a list of confirmed errors of the same kind, and check them
one by one.

In~\cite{horvath2019categorization}, authors collected a list of commits from the version
control history of the LLVM codebase that were fixing lifetime-related bugs. Their reason
was to test a different kind of toolset solving the same problem: compiler warnings. They
would wind back the code to a state before the fix, and see if the warnings find the problem.

Fortunately for us, many of the historical bugs listed were caused by \texttt{StringRef}s,
LLVM's own \texttt{string\_view} implementation. Although we do not support this special
class in our checkers, its semantics are almost identical to those of \texttt{std::string\_view},
and it is relatively straightforward to change the \texttt{StringRef}s involved in the bug
to \texttt{std::string\_view}s.

Out of the 22 cases listed, 10 were bugs caused by the incorrect use of \texttt{StringRef}.
After rewriting the cases to \texttt{std::string\_view}, our checkers found all of them.
In Table~\ref{table}, we list the 10 LLVM commits examined. (Note that the LLVM project
has moved to a new repository\footnote{https://github.com/llvm/llvm-project} since 
\cite{horvath2019categorization} was published. We
have updated the commit hashes to point to the current repository.)

\begin{table}[h]
\footnotesize
\centering
\label{table}
\begin{tabular}{|c|}
\hline
fec6b1afe66ccfeff0c80c9ed0aea1f6daaf5832   \\ \hline
7054b49e0193de6557676319fb1af9b59cc6b333   \\ \hline
ea8d370d2a422b0d017916d0507484be1d8808ba   \\ \hline
3eb0edde78e57422a0e1856d1085a218368e4635   \\ \hline
33cd6dcf87295e4165fe9fca56b078eb686d044a    \\ \hline
6465d4f7a36077df16188d3d9d512f951333851a    \\ \hline
dadfc1c144389b6917fa127d1e178a055ef70376    \\ \hline
30a243268209377344eda599277ed2475dff9dd6    \\ \hline
9eef7423081e6427086a9243b380a9e254067b05    \\ \hline
45ed4f954265175849efdd0d45940a64395b44d0    \\
\hline
\end{tabular}
\caption{A list of LLVM commits fixing \texttt{StringRef} related memory errors. 
Our checkers found the errors after rewriting them to use \texttt{std::string\_view}.}
\end{table}

An excerpt from one of the old errors edited to use \texttt{std::string\_view} can 
be seen in Listing~\ref{lst:gnu}. On line 5, the \texttt{std::string\_view} is created
from a temporary \texttt{std::string} returned by the \texttt{StringRef::str} method.
Because the string is destroyed at the end of the full-expression, the view is already
dangling on line 6. We decided not to consider the \texttt{empty} method as an erroneous
use, so our checkers only warn on line 8, when the \texttt{find\_first\_not\_of} method
is called.

\begin{lstlisting}[caption={A highly simplified version of an old \texttt{StringRef} bug in LLVM,
turned into a \texttt{std::string\_view} bug, and found by our analysis.},
label={lst:gnu},basicstyle=\ttfamily\footnotesize]
void Generic_GCC::GCCVersion::Parse(StringRef VersionText) {
  std::pair<StringRef, StringRef> First = VersionText.split('.');
  std::pair<StringRef, StringRef> Second = First.second.split('.');
  
  std::string_view PatchText = Second.second.str();
  if (!PatchText.empty()) {
    if (size_t EndNumber = PatchText.find_first_not_of("0123456789")) {
      // ...
    }
  }
}
\end{lstlisting}

%%%%%%
\section{Future work}

One of our most important plans for the future is to upstream our work to the
official Clang repository, to enable it to reach a wide audience.
Clang, part of the LLVM Compiler Infrastructure Project, is one of the two most 
widely used open-source C++ compilers in the world, together with GCC. Our lab
has a history of contributions to the Clang Static Analyzer, which gives us
confidence that the upstreaming will be successful. However, it is a long process
that may take months, or even years. This is why we decided to publish our
working draft (see Section~\ref{how-to}).

We gave details in Section~\ref{eval} about the state of adoption of the 
standard version of the \texttt{string\_view} data structure. We expect more
and more C++ projects to migrate to the latest standards and rewrite their
codebases to use the new, standard version of the concept. In the meantime,
however, we plan to extend our analysis to work for certain popular non-standard
implementations. Our feeling is that supporting Google's \texttt{absl::string\_view},
LLVM's \texttt{llvm::StringRef}, or the Boost libraries' \texttt{boost::string\_ref}
would benefit a large portion of the C++ community.

%%%%%%
\section{Conclusion}

The majority of the security vulnerabilities found in the C/C++ codebases of tech 
companies like Microsoft and Google stem from memory errors.
With this paper, we join the battle against lifetime-related problems in C++ by
providing a set of bug finding tools that find incorrect uses of 
the popular \texttt{string\_view} data structure early in the development process.

We showed that \texttt{string\_view} is a fast and cheap means of string processing,
and described the memory errors it may cause. We detailed the logic behind the
static analysis tools We implemented as part of the popular Clang compiler, to
ensure that it reaches a  wide audience. Our feature is not yet available in the
official repository, but we plan to upstream it in the future.

The C++ community has not yet fully adopted the standard \texttt{string\_view} class
introduced in 2017. Most of the projects that do extensive string processing use
earlier implementations of the same concept, which constrained the evaluation of our
tools. We did, however, run our analysis on a historical list of lifetime errors
in the LLVM codebase, and found that our tools would have found those errors. We 
also ran our analysis on a list of widely used open-source projects to confirm that 
it does not produce bogus reports.

\section*{Acknowledgement} The publication of this paper is supported by the 
European Union, co-financed by the European Social Fund 
(EFOP-3.6.3-VEKOP-16-2017-00002).

\bibliographystyle{plain}
\bibliography{bib}

\end{document}